\documentclass[useAMS,usenatbib]{mn2e}

\usepackage{gensymb}
\usepackage{graphicx}	
\usepackage{amsmath}	
\usepackage{amssymb}

 \title[Isotropy analyses of the Planck convergence map]
          {\mbox{Isotropy analyses of the Planck convergence map}}
\author[G. A. Marques, C. P. Novaes, A.  Bernui, and I. S. Ferreira]{
G. A. Marques$^{1}$\thanks{e-mail: gabrielamarques@on.br} 
C. P. Novaes$^{1}$\thanks{e-mail: camilapnovaes@gmail.com} 
A. Bernui$^{1}$\thanks{e-mail: bernui@on.br} 
and I. S. Ferreira$^{2}$\thanks{e-mail: ivan@fis.unb.br} \\
$^{1}$Observat\'orio Nacional, Rua General Jos\'e Cristino 77, 
          S\~ao Crist\'ov\~ao, 20921-400 Rio de Janeiro, RJ, Brazil \\
$^{2}$Instituto de F\'{\i}sica, Universidade de Bras\'{\i}lia, 
Campus Universit\'ario Darcy Ribeiro, Asa Norte, 70919-970, Bras\'{\i}lia, DF, Brazil 
}
\date{Accepted 2017 August 28. Received 2017 August 23; in original form 2017 January 04}
\pubyear{2017}
\begin{document}
\label{firstpage}
\pagerange{\pageref{firstpage}--\pageref{lastpage}}
\maketitle
\begin{abstract}
The presence of matter in the path of relic photons causes distortions in the angular pattern of the cosmic microwave background (CMB) temperature fluctuations, modifying their properties in a slight but measurable way. 
Recently, the Planck Collaboration released the estimated \textit{convergence map}, an 
integrated measure of the large-scale matter distribution that produced the weak gravitational lensing (WL) phenomenon observed in Planck CMB data. 
We perform exhaustive analyses of this convergence map calculating the variance in small and large regions of the sky, but excluding the area masked due to galactic contaminations, and compare them with the features expected in the set of simulated convergence maps, also released by the Planck collaboration. 
Our goal is to search for sky directions or regions where the WL imprints anomalous signatures to the variance estimator revealed through a $\chi^2$ analyses at a statistically significant level. 
In the local analysis of the Planck convergence map we identified 8 patches of the sky in disagreement, in more than 2$\sigma$, with what is observed in the  average of the simulations. 
In contrast, in the large regions analysis we found no statistically significant discrepancies, but, interestingly, the regions with the highest $\chi^2$ values are surrounding the ecliptic poles.
Thus, our results show a good agreement with the features expected by the $\Lambda$CDM concordance model, as given by the simulations.
Yet, the outliers regions found here could suggest that the data still contain residual contamination, like noise, due to over- or under-estimation of systematic effects in the simulation data set.

\end{abstract}


\section{Introduction}

Observations of the temperature and polarization of the cosmic microwave background (CMB) data 
provide crucial information about the early universe (see, e.g., 
\cite{durrer2001theory, hu2000weak}). 
However, these data also encode important signatures from secondary processes, generated after 
the last scattering surface (LSS), which contribute to blur the primordial features of the 
CMB radiation~\citep{weiland2010seven,serra2008impact,aghanim2008secondary}. 
One of these secondary processes, the weak gravitational lensing (WL), consists in deflections 
of the CMB photons coming from the LSS by the inhomogeneous distribution of matter along our 
line of sight \citep{blanchard1987gravitational,cole1989gravitational,linder1990analysis}. 
The main mensurable effect resulting from this phenomenon is the smoothing of the acoustic 
peaks in the CMB angular power spectrum, reducing the contrast of the peaks and troughs 
\citep{Lewis}. 
Since these data is used to assert the parameters of competing cosmological models, it is 
important to correctly discriminate this WL effect on CMB data.

The WL on CMB photons has been investigated and measured by various methods and diverse 
experiments in the recent past 
\citep{hirata2004cross,smith2007d,hirata2008co,smidt2011constraint,das2011detection,keisler2011me,
feng2012mea,feng2012recon,van2012measurement,story2013measurement,das2014atacama}. 
However, only with the observations of the Planck satellite, that improved the detector 
sensitivity getting reliable knowledge of the noise, the dominant signal at small angular 
scales~\citep{ade2014planck,ade2015planck}, it was possible the robust reconstruction of the 
{\it lensing potential map}, LPM, which consist in the sky projection of the 
gravitational potential causing the WL of the CMB photons.

The convergence map is related to the gravitational lensing potential via the 2-dimensional 
(2D) Poisson equation and can be interpreted as the sky projection of the surface density of the 
(baryonic plus dark) matter structures that CMB photons encounter in their path between the LSS 
and us. 
Because different cosmological models predict diverse scenarios of how the matter clumps, the convergence map 
features can be beneficial to test these possibilities, or even be used to improve parameter 
constraints within the concordance cosmology, $\Lambda$CDM, particularly for parameters affecting 
the structure formation in the universe~\citep{adam2015planck}.
As a matter of fact, the Cosmological Principle asserts that the matter and radiation are 
isotropically and homogeneously distributed around us at large scales. 
For this, many efforts have been done to test this property of the universe using diverse 
cosmological probes, like the CMB data~\citep{ghosh2016dipole, Schwarz2015cma, bernui2014north}, 
the gamma-ray bursts \citep{tarnopolski2015testing, bernui2008large,ukwatta2016inve}, the angular distribution of galaxies 
\citep{bengaly2017there,alonso2015homogeneity, tiwari2016revisiting} and galaxy clusters 
\citep{bengaly2015probing}, for instance.

In this sense, our main motivation here is to test how isotropically distributed is the matter in the universe 
according to information contained in the convergence map and, moreover, if this result is in agreement with what is 
expected in the concordance cosmology, $\Lambda$CDM. 
Subsequently, we perform analyses upon the Planck lensing data set, that is, the simulated and the 
estimated convergence map, evaluating how the empirical variance (following a procedure described below) is 
distributed in the whole sky. 

This paper is organized as follows: 
section~\ref{sec:lenspot} presents a brief review about the WL effect upon the CMB photons. 
Details about the weak lensing products released by the Planck collaboration, besides the 
analyses performed upon them, are described in section~\ref{sec:methodology}. 
The results and discussions are presented in section~\ref{sec:results} and the conclusions are 
addressed in section~\ref{sec:conclusion}.

\section{Weak Gravitational lensing of the CMB}
\label{sec:lenspot}
 
The deviation of the CMB photons path by potential gradients along the line of sight from the LSS until 
they reach us can be described by a deflection vector field, expressed as
$\alpha(\hat{\Omega}) \equiv \mathbf{\nabla}\psi(\hat{\Omega})$, where $\mathbf{\nabla}$ is the 2D 
gradient operator on the sphere and the $\psi(\hat{\Omega})$ parameter corresponds to the lensing 
potential in the direction $\hat{\Omega}$. 
The deflection also can be related in terms of the convergence term, $\kappa= -\frac{1}{2}\nabla \cdot \alpha$. 

The lensing potential $\psi$ is the projection along the line-of-sight of the gravitational potential, $\Psi(r,\hat{\Omega})$, at conformal distance $r$, that is,
\begin{equation} 
\centering
\psi(\hat{\Omega})= -2\int^{r_{0}}_{0} dr \frac{d_{A}(r_{0}-r)}{d_{A}(r)d_{A}(r_{0})}\Psi(r,\hat{\Omega}) \, ,
\label{eq:potlens}
\end{equation}
where $r_{0}$ is the conformal distance to the LSS and $d_{A}$ is the comoving angular diameter distance. Since the convergence is related to the lensing potential via the 2D Poisson equation, $\kappa=-\frac{1}{2}\nabla^2\psi$, it can be interpreted as a (projected) surface density. 
 
Lensing, in General Relativity, is an achromatic effect, that is, it remaps the CMB fluctuations without changing the frequency dependence and number of photons. 
The redistribution of the CMB temperature fluctuations on the sky through angular deflections can be 
expressed by $\tilde{\Theta}(\hat{\Omega})=\Theta(\hat{\Omega} + \mathbf{\alpha})$, where 
$\Theta(\hat{\Omega}) = \Delta T(\hat{\Omega})/T_0$, with $T_0 = 2.725 K$~\citep{kogut1996microwave}. 
The deflections of photon's paths generate distortions in the angular pattern of the primary 
temperature fluctuations so that, expanding the above expression in a Taylor series one can write as approximation~\citep{Lewis} 
\begin{equation}
\begin{split}
\tilde{\Theta}(\hat{\Omega})=\Theta(\hat{\Omega}+ \mathbf{\nabla}\psi) \simeq 
\Theta(\hat{\Omega}) + \nabla^{a} \Theta(\hat{\Omega})\nabla_{a} \psi(\hat{\Omega})+ \\
\frac{1}{2}\nabla^{a}\psi(\hat{\Omega})\nabla^{b}\psi(\hat{\Omega})\nabla_{a}\nabla_{b}\Theta(\hat{\Omega})+\ldots
\label{eq:expansao}
\end{split}
\end{equation} 

The remapping of the CMB fluctuations changes in few percent the angular power spectrum, 
$C_{l}^{\Theta}$, mainly at small scales \citep{hu2000weak, 2010/lewis}. 
From the Taylor approximation in equation (\ref{eq:expansao}) the lensed power spectrum 
$\tilde{C}_{l}^{\Theta}$ can be written as 
\begin{eqnarray}
\tilde{C}_{l}^{\Theta} \,\simeq\, 
C_{l}^{\Theta} &+& \int\frac{d^2\mathbf{l'}}{(2\pi)^2}[\mathbf{l'}\cdot (\mathbf{l-l'})]^2 
C_{|\mathbf{l-l'}|}^{\psi}C_{l'}^{\Theta} \nonumber \\ 
&-& C_{l}^{\Theta}\int\frac{d^2\mathbf{l'}}{(2\pi)^2}(\mathbf{l\cdot l'})^2C_{l'}^{\psi}+\ldots 
\end{eqnarray}
That is, the lensed power spectrum depends on the convolution of the unlensed temperature power 
spectrum with the lensing potential power spectrum, $C_{l}^{\psi}$. 
This convolution transfers power from large to small scales making the resulting lensed power spectrum dominant over the primary one at $l \geq 5000$ \citep{hanson2010weak}. 

In addition, the effect of WL also mixes the E-mode polarization into B-mode polarization 
\citep{bmode} and produces a distinctive small-scale non-Gaussian trispectrum (four-point 
correlation function), with smaller signals in the higher even n-point functions 
\citep{hanson2009cmb}. 
The WL effect is a secondary effect that changes slightly the primary CMB anisotropy features, 
however provides important cosmological constraints complementary to the CMB data 
\citep{ade2015planck} and to the large scale structure 
\citep{liu2015cross,liu2016origin,liu2016cmb,kirk2016cross,singh2017cross}. 

In the above paragraphs we summarized the expected angular distribution features of the CMB 
lensed photons, with no indications for preferred directions. 
For this, it is worth to test the statistical isotropy of the Planck WL products, at small 
and large regions, 
to investigate whether other effects (like foregrounds, masking procedure, 
etc.) may have introduced anomalous, and unaccounted, deviations from isotropy in the data.
 

\begin{figure}
\centering
\hspace*{-1.3cm}                                                      
\includegraphics[scale=0.4]{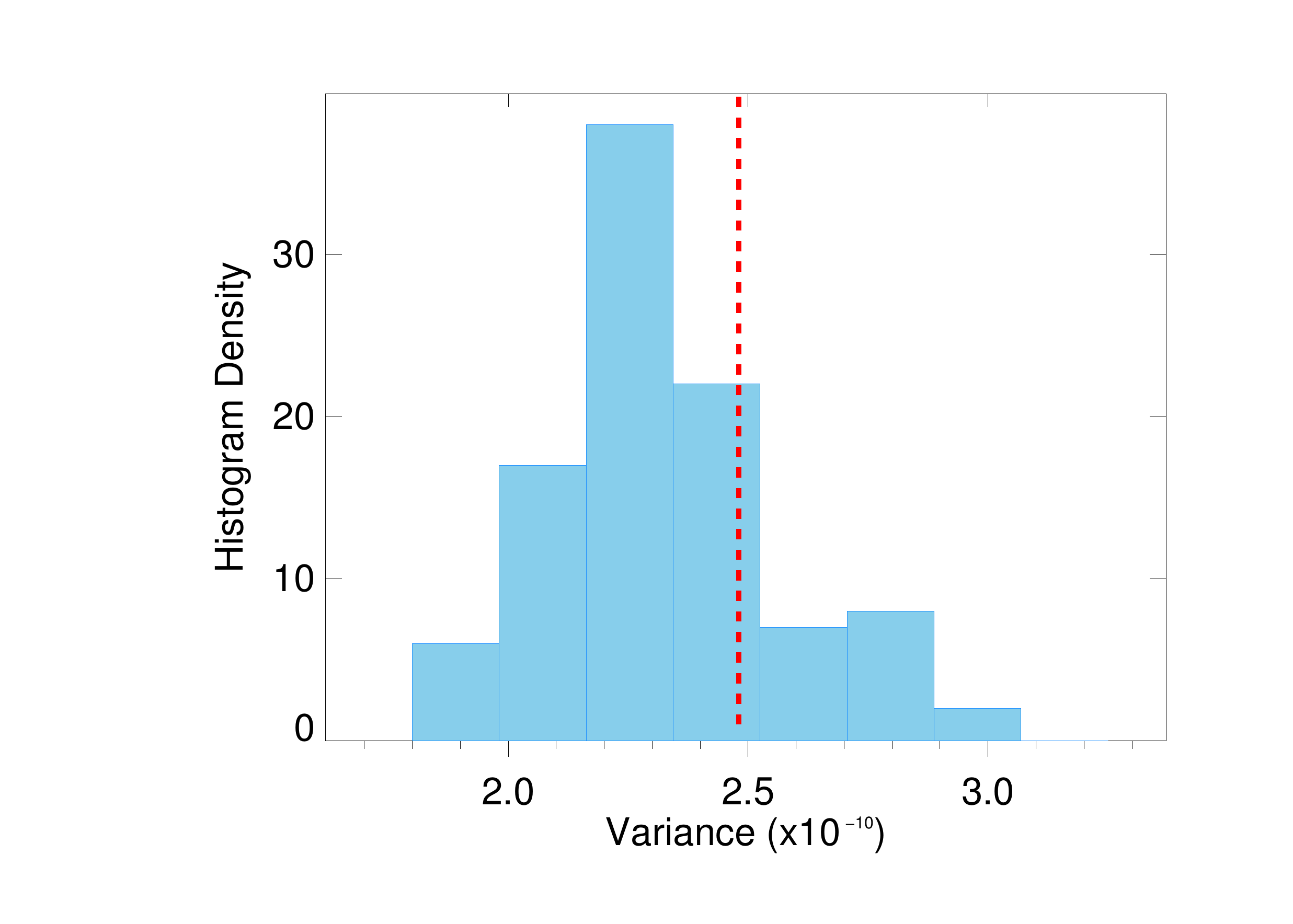}
\vspace{-1cm}
\caption{Histogram of the variance for the simulated Planck dataset, $\sigma^2_{(sims)}$,
calculated excluding the WL mask region. 
The vertical red dashed line corresponds to the variance of the Planck convergence map  $\sigma^2_{(planck)}$.
} 
\label{fig:histo_sky}
\end{figure}
 

\section{Data and methodology}
\label{sec:methodology}

\subsection{The Planck convergence map}
In 2015, the \textit{Planck} collaboration performed a precise measurement of CMB lensing power 
spectrum, with a 40$\sigma$ detection \citep{ade2015planck}. 
To reconstruct the CMB lensing potential map the Planck team used quadratic 
estimators~\citep{okamoto2003cosmic} that make use of the features induced by the lensing process, 
that is, the diverse correlations of the CMB temperature (T) and polarization (E and B) modes. 
A combination of these estimators in a minimum-variance (MV) estimator was used to reconstruct the 
CMB lensing potential, $\hat{\psi}^{\rm MV}$.
The CMB dataset used as input to the MV lensing estimator was the foreground-cleaned map obtained 
applying the \texttt{SMICA} code to the raw Planck 2015 full-mission frequency 
maps~\citep{adam2015planck}. 
The Planck team provided the multipole expansion coefficients of the estimated lensing 
convergence map, $\kappa$, which can be written as a function of the lensing potential spherical 
harmonic coefficients as 
\begin{equation}
\kappa_{LM}^{\rm MV} \,=\, \frac{L(L+1)}{2}\hat{\psi}_{LM}^{\rm MV} \, , 
\end{equation}
where $L$ and $M$ are the multipole components indices for the lensing reconstructed map. 

The noise reduction of the maps is performed by applying a Wiener filter in spherical harmonics \citep{bobin2012cmb} 
\begin{equation}
\hat{\psi}_{LM}^{\rm WF} \,=\, 
\bigg(\frac{C_{L}^{\psi,fid}}{C_{L}^{\psi,fid}+N_{L}^{\psi}}\bigg)\hat{\psi}_{LM}^{\rm MV} \, ,
\label{eq:wigner}
\end{equation}
where $C_{L}^{\psi,fid}$ is the lensing potential power spectrum according to the fiducial cosmological 
model and $N_{L}^{\psi}$ is the noise power spectrum of the $\psi$ reconstruction.
The lensing convergence and the noise on $\kappa$ have a much whiter power spectrum on large scales while the estimated lensing potential has a very red power spectrum \citep{adam2015planck}. 
Cutting the maps with red power spectrum in small portions can cause leakage issues. 
For this reason, we use in our analysis the Wiener-filtered convergence map, $\hat{\kappa}^{\rm WF}$, 
instead of the lensing potential map, as described in the next section. 
It was produced considering the resolution $N_{side} = 2048$, using the \texttt{HEALPix} 
(Hierarchical Equal Area iso-Latitude Pixelization) pixelization grid \citep{gorski2002hea}. 
Our analyses are performed using the corresponding confidence mask (hereafter called WL mask), 
also released among the Planck lensing products. 
This mask joins of the Galaxy mask, that removes 
$f_{sky} \simeq 0.302$, the point-sources mask that removes additional $f_{sky}\simeq 0.07$  
\citep{ade2014planck} and the \texttt{SMICA} specific temperature and polarization mask 
\citep{adam2015planck}. 
The combination of these masks leaves a total unmasked sky fraction of $f_{sky} \simeq 0.673$.

\subsection{Planck Lensing simulations} 
The performance of the quadratic estimators used to reconstruct the lensing maps is 
affected by non-lensing sources such as foreground residuals and inhomogeneous 
instrumental noise, besides the effects of beam asymmetry and masking. 
To account for these biases and to characterize the variance limits of the reconstructed 
convergence map $\hat{\kappa}^{\rm WF}$, we used a set of 100 realistic realizations of 
this observable, also released by the Planck team.
These simulations are based on the Full Focal Plane 8 (FFP8) Monte Carlo realizations 
which is a set of maps that incorporate the dominant instrumental (detector beam, bandpass 
and correlated noise properties), scanning (pointing and flags) and data analysis (map-
making algorithm and implementation) effects as described in~\cite{fullfocal}. The FFP8 simulations do not include galactic foregrounds contributions. In the lensing analysis, residual foregrounds in the SMICA maps are modelled by adding a small level of statistically-isotropic Gaussian noise, with an appropriate angular power spectrum, to the simulated frequency maps after combination with the SMICA weights. It is worth mentioning that the noise distribution in the data is highly anisotropic and any mis-modelling of the simulated noise may have implications for the current analysis.

All the non-lensing effects included in the reconstruction process make the $\hat{\kappa}^{\rm WF}$ data 
map highly non-isotropic. 
To study how the variance of the WL is distributed on the sky, beyond the contribution of those effects, 
we considered in our analysis the MV convergence simulated dataset after applying a Wiener-filter to them using equation~(\ref{eq:wigner}), similarly to the procedure described by the Planck Collaboration to obtain the estimated convergence map.

The reconstruction of the data is band-limited to $8\leq L \leq 2048$ due the instability upon low order 
multipoles \citep{adam2015planck}. 
For this reason, we verify that the simulated maps contain the same multipole range as well the 
same pixelization resolution as the Planck data. 

Moreover, to confirm that the Planck simulated maps have their variances in agreement with that 
of the reconstructed $\hat{\kappa}^{\rm WF}$ map, we plot in figure~\ref{fig:histo_sky} the histogram 
of their empirical variance upon almost full-sky, that is, cutting only the area enclosed 
by the WL mask. 
The vertical dashed line represent the variance of the estimated Planck convergence map, 
$\hat{\kappa}^{\rm WF}$. 
The average value from the 100 simulated maps is in complete agreement with the variance of the data map, that is, they have the same value up to 1$\sigma$. 
 
\subsection{Analyses of the Planck convergence map} 
 
The main goal of our analyses is to test how isotropic is the Planck convergence map. The variance provides a simple way to look for statistically significant deviations compared to the expected by the simulated convergence maps in different regions of the sky. 

In this sense, we choose to perform our analysis of the convergence maps using two 
complementary approaches: 
(1)  performing local analyses calculating the lensing variance in non-overlapping patches of 
the sky, and 
(2) through a hemispherical scan of the sky, in order to determine if there is a preferred direction, statistically significant in comparison to the expected from the simulations.
Each method is described in detail below. 
\subsubsection{Local analysis}
We define the sky regions where to carry out the local analysis as being the pixels 
corresponding to a \texttt{HEALPix} resolution of $N_{side} = 4$, that is, 192 pixels of equal 
area $\sim (14.7 ~{\rm deg})^2$. 
Each one of these big pixels in the small resolution (hereafter called \textit{patches}) contains $\sim$ 262,144 
small pixels in the high resolution, $N_{side} = 2048$. 
 
Since we used the WL mask to remove Galactic and extragalactic contamination, the number of 
valid pixels in each region, after the application of the mask, varies from one region to another.
Then, we established the criterion of discarding the patches whose number of valid pixels is less 
than 80$\%$ of the total. As a result, a total of 116 patches were selected for analysis.
 
Aware of how the WL effects are distributed on the sky, our analysis uses the variance of the lensing 
convergence as the statistical estimator to measure the directional amplitude of this effect. 
For each of the $p$-patches ($p \in [1,116]$), we calculate this quantity in all the simulated, $\sigma^2_{p(Sims)}$, and Planck estimated, $\sigma^2_{p(Planck)}$, convergence maps.

\begin{figure}
\centering
\hspace{-0.3cm}
	\includegraphics[scale=0.33, angle=90]{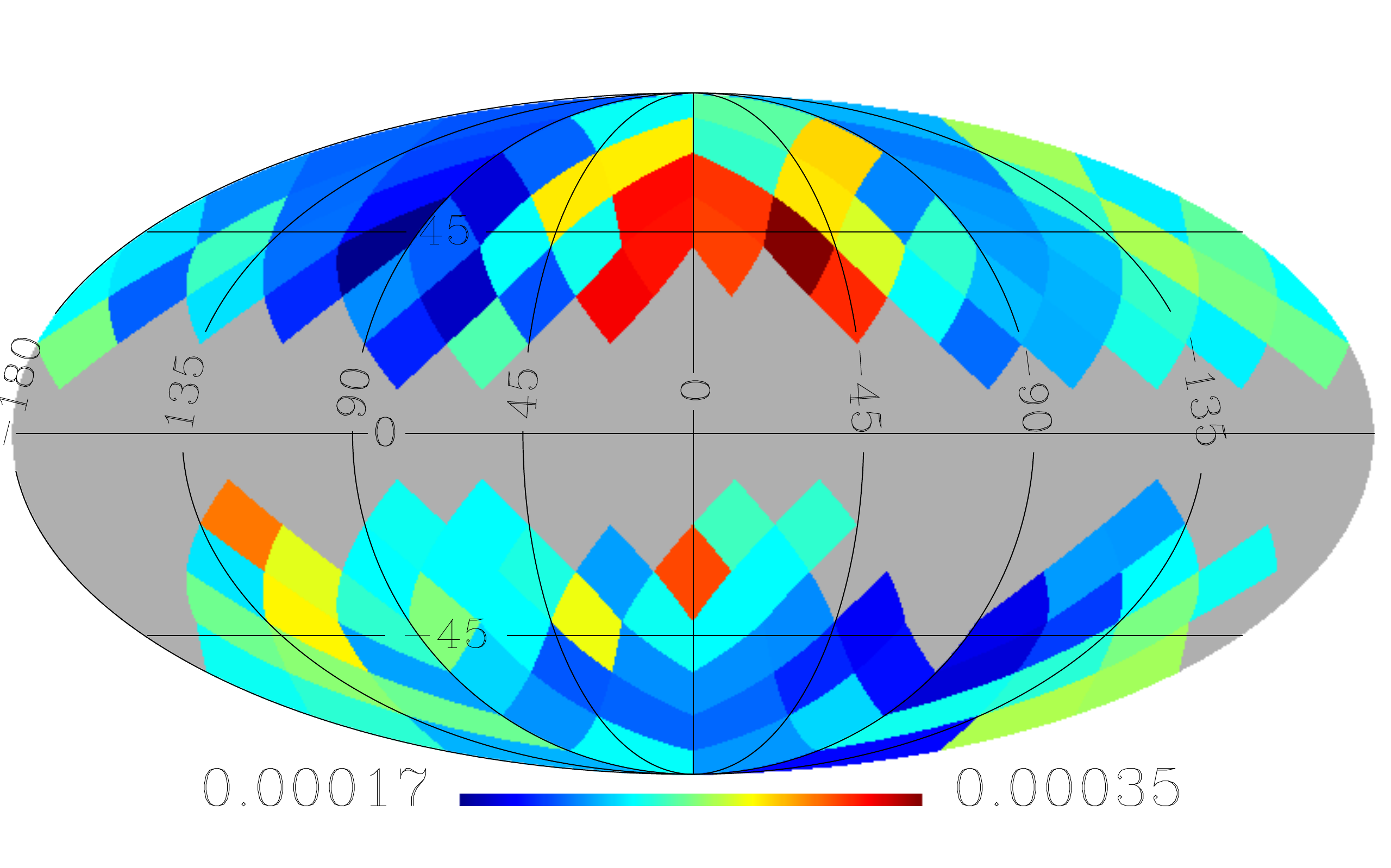}
\vspace{-0.7cm}
	\caption{Mollweide projection of the variance $\sigma^2_{p(Planck)}$ calculated in patches of the 
	estimated Planck convergence map, $\hat{\kappa}^{WF}$ (see text for details). 
	The region excluded from the analysis by the WL mask is presented in grey.} 
	\label{fig:variance_plan}
\end{figure}

\begin{figure}
\centering
\hspace{-0.3cm}
	\includegraphics[scale=0.33, angle=90]{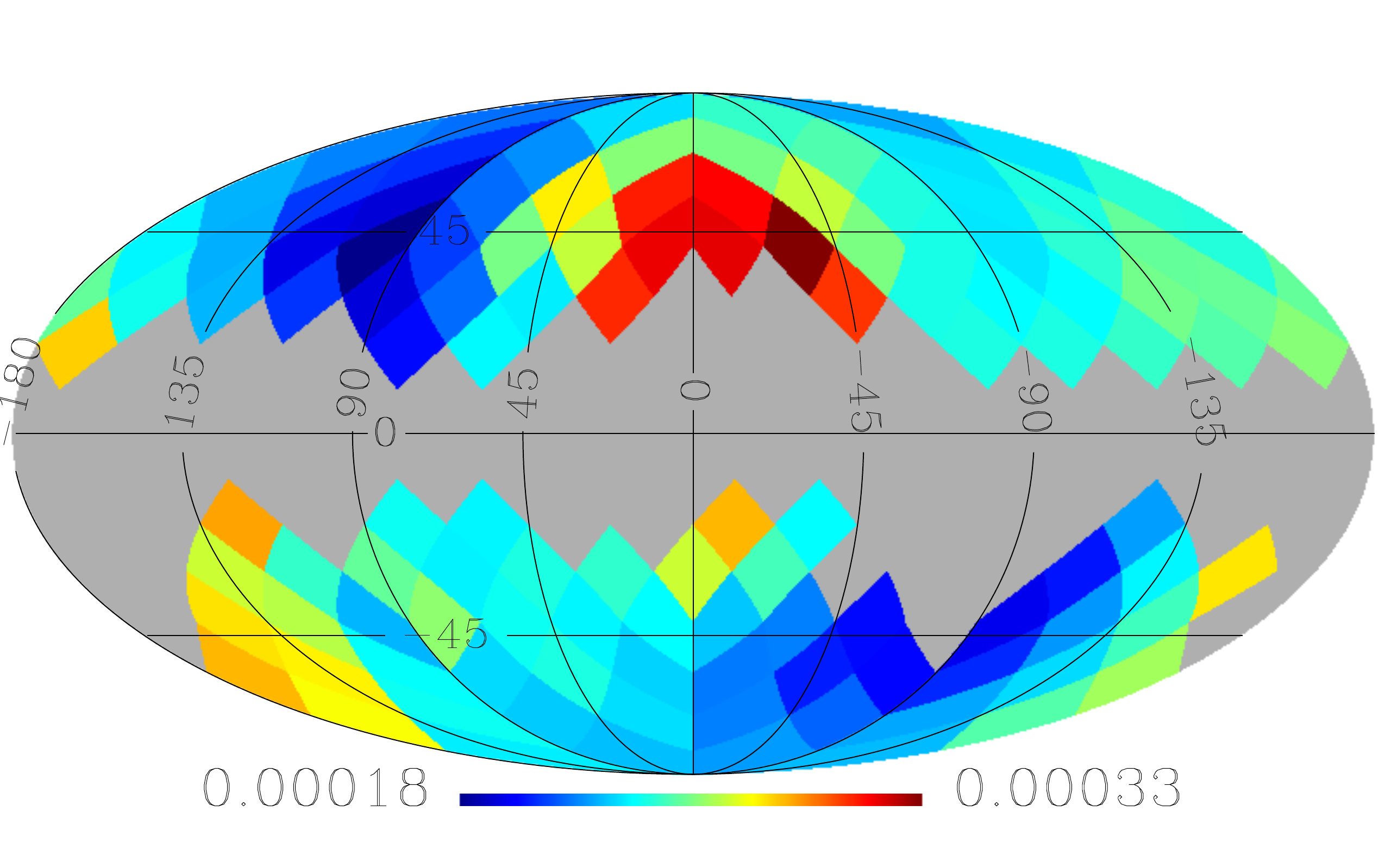}
\vspace{-0.7cm}
	\caption{Mollweide projection of the mean variance $\langle{\sigma}^2_{p(Sims)} \rangle$, 
	similar to the analysis presented in figure \ref{fig:variance_plan}, but averaging upon the 
	100 simulated convergence maps (see text for details).} 
	\label{fig:variance_sim}
\end{figure}

The figure \ref{fig:variance_plan} shows, in color scale, a Mollweide projection of the $\sigma^2_{p(Planck)}$ for 
each of those patches, and the figure \ref{fig:variance_sim} presents the mean variance, 
$\langle{\sigma}^2_{p(Sims)}\rangle$, calculated from the simulated dataset.


\subsubsection{Hemispherical analysis}
In contrast, the hemispherical analysis applies the empirical variance estimator upon large regions of the sky. The centres of the hemispheres are defined by pixelizing the celestial sphere according to 
the resolution parameter $N_{side}=4$. 
This provides a total of 192 coordinates centres. 
Once more, we used the WL mask to remove the residual contamination region. 
Since in the current case we are considering regions corresponding to hemispheres, which 
encompasses large areas of the sky, the percentage of valid pixels variate slightly in each of them. 

Similarly to the local analysis, we calculate the variance in each one of the $h$-hemispheres, 
$h \in [1,192]$, upon all the simulated, ${\sigma}^2_{h(Sims)}$, and 
estimated, $\sigma^2_{h(Planck)}$, Wiener-filtered convergence maps. 
The Mollweide projection of the hemispherical variance of the reconstructed Planck convergence map, 
${\sigma}_{h(Planck)}^2$, and the corresponding average from the simulated dataset, 
$\langle{\sigma}^2_{h(Sims)}\rangle$, are shown in 
figures~\ref{fig:Hemispherical_planck} and~\ref{fig:Hemispherical_simul}, respectively.

\section{Results and Discussion}
\label{sec:results}  
\begin{figure}
	\centering  
	\hspace{-0.3cm}                                                       
	\includegraphics[scale=0.33,angle=90]{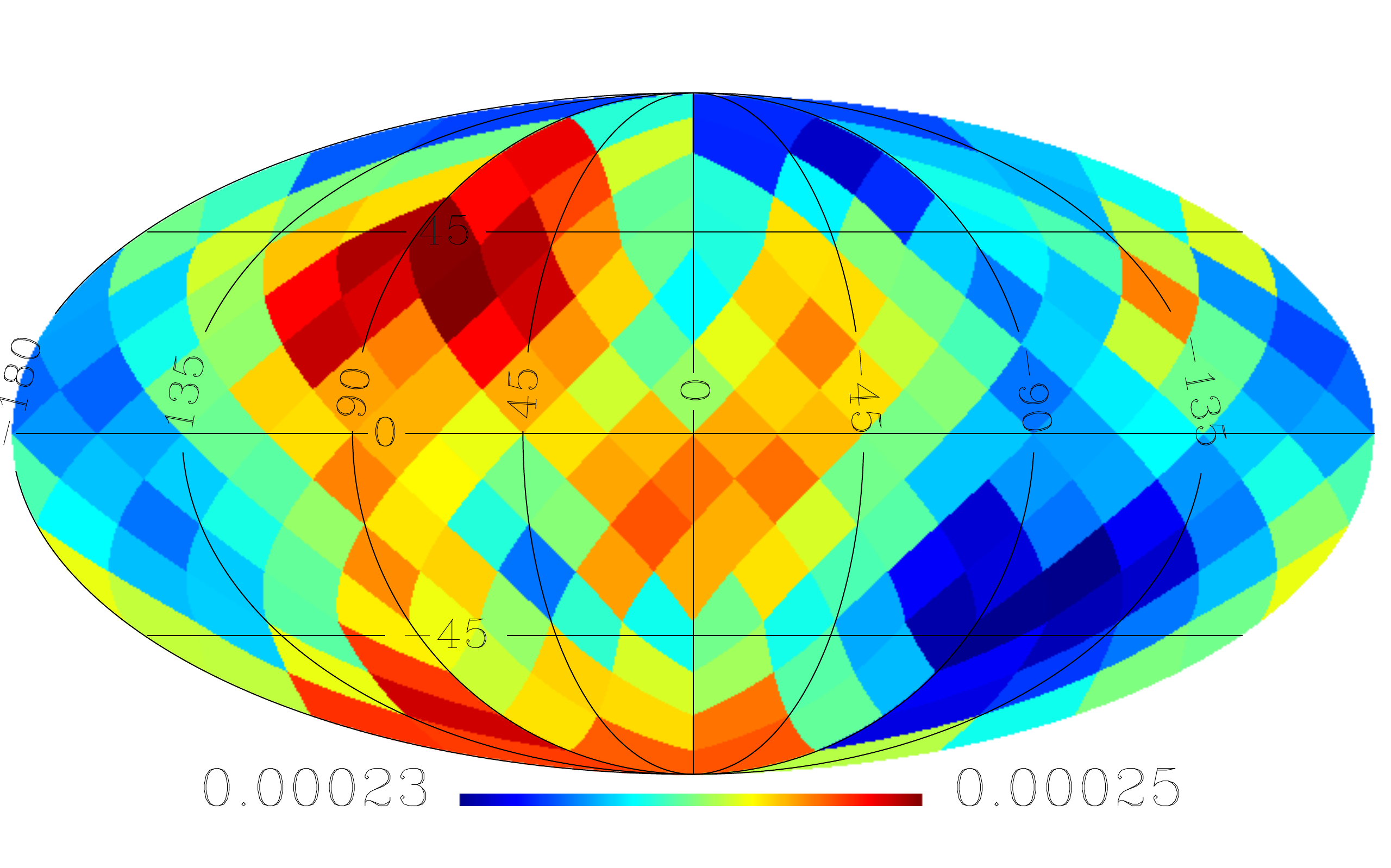}
	\vspace{-0.7cm}
	\caption{Mollweide projection of the variance, $\sigma^2_{h(Planck)}$, obtained performing the hemispherical analysis of the Planck convergence map, $\hat{\kappa}^{WF}$, and projected at the 192 centres, as explained in the text.
	} 
	\label{fig:Hemispherical_planck}
\end{figure}

\begin{figure}
	\centering  
	\hspace{-0.3cm}                                                       
	\includegraphics[scale=0.33,angle=90]{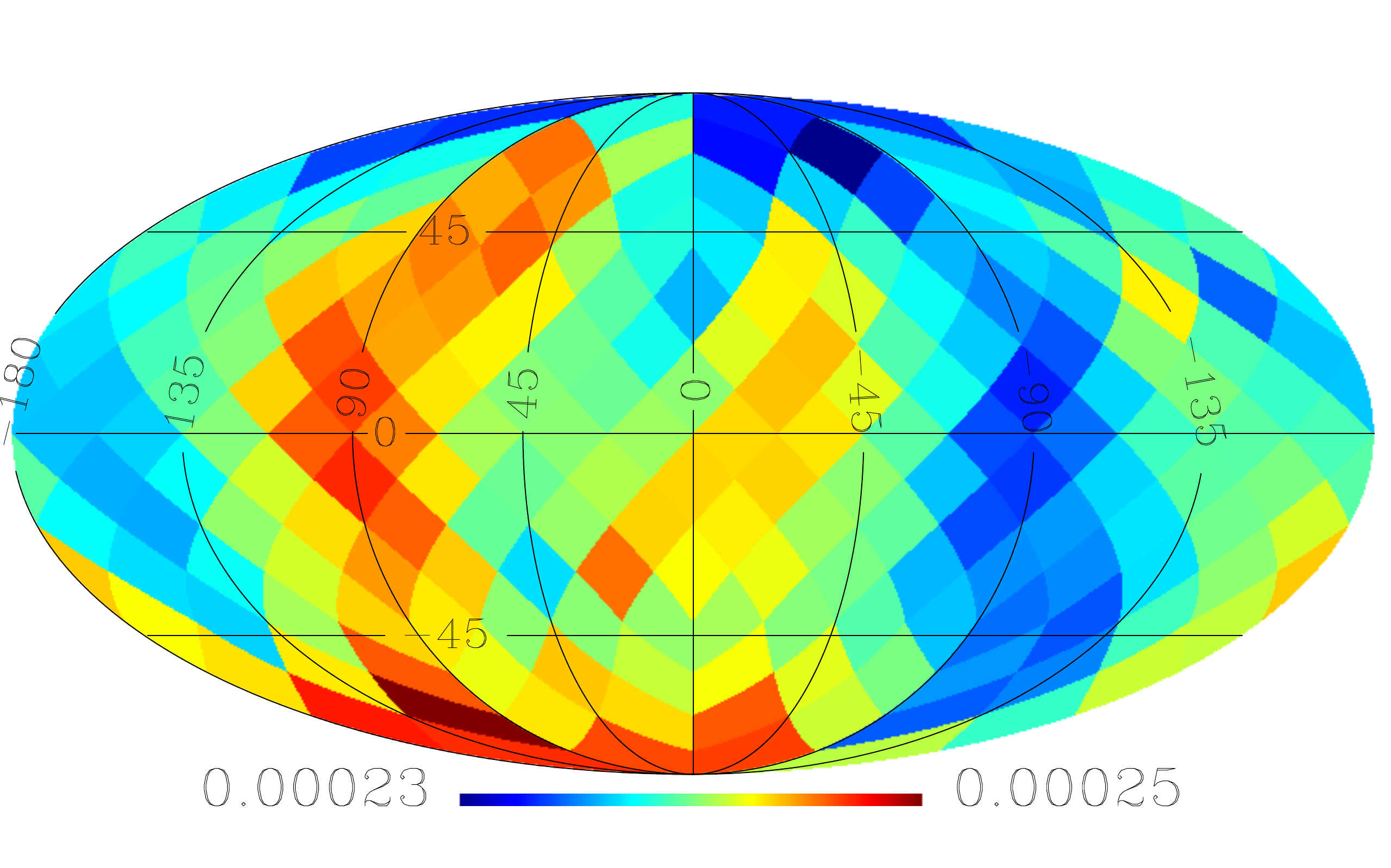}
	\vspace{-0.7cm}
	\caption{Mollweide projection of the $\langle{\sigma^2_{h(Sims)}}\rangle$ resulting from a hemispherical analysis, similar to the analysis presented in figure \ref{fig:Hemispherical_planck}, but averaging upon the values obtained from the 100 Planck simulated convergence maps (see text for details). 
	} 
	\label{fig:Hemispherical_simul}
\end{figure}

\begin{figure}
\begin{minipage}{.6\textwidth}
\hspace{0.35cm}
\includegraphics[width=0.7\textwidth]{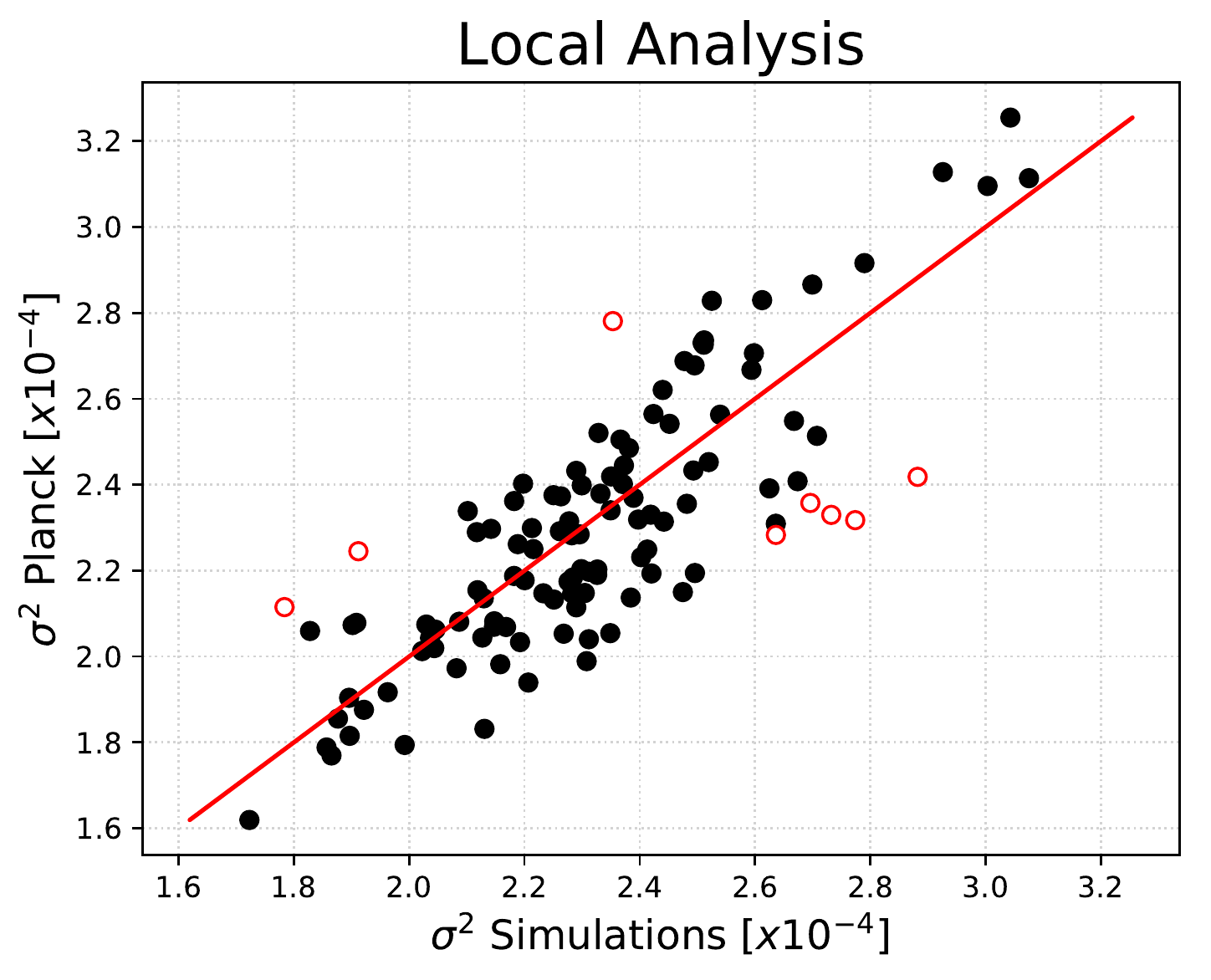}\\
\end{minipage}
\begin{minipage}{.6\textwidth}
\vspace{-0cm}
\hspace{0.35cm}
\includegraphics[width=0.7\textwidth]{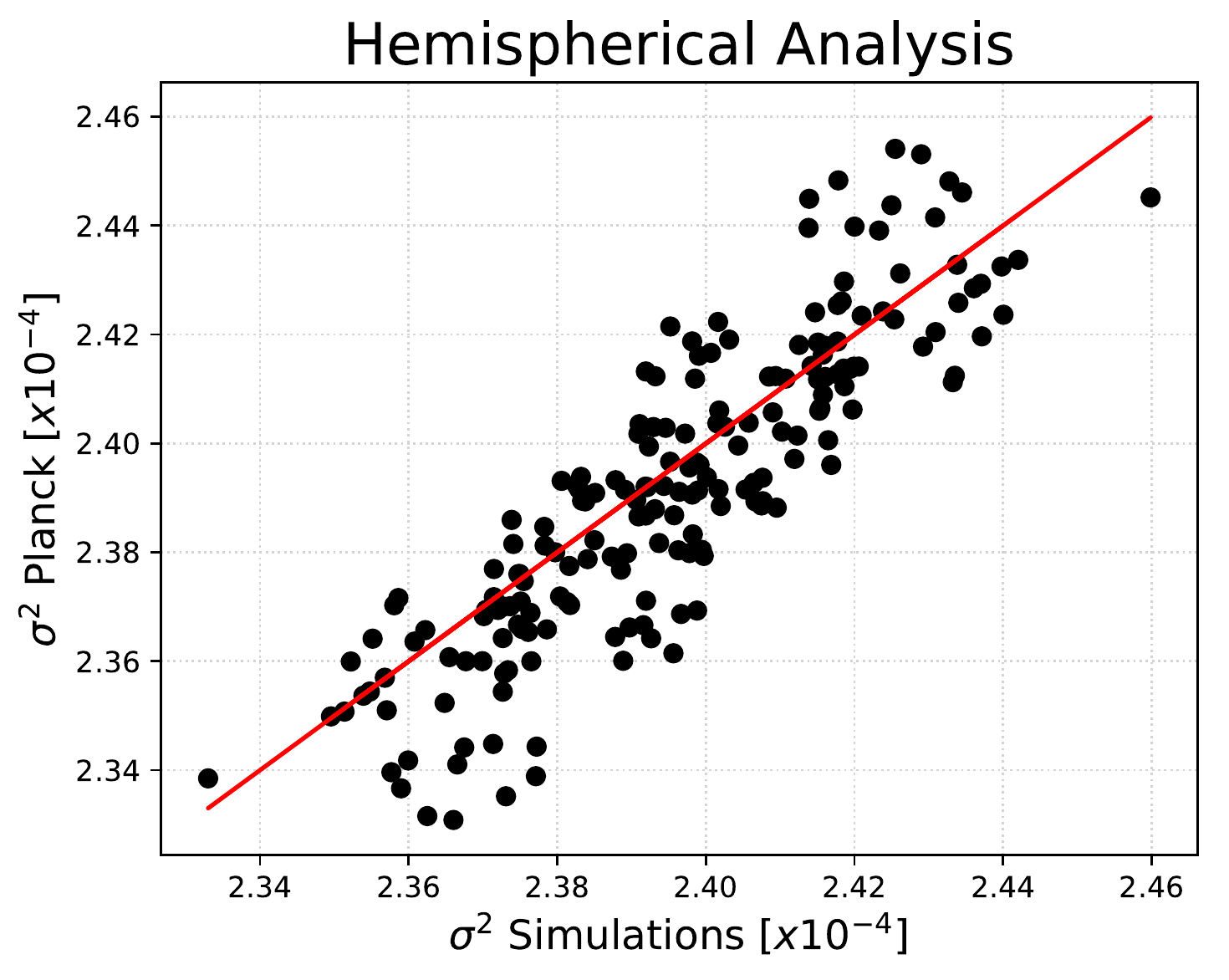} 
\end{minipage}

\caption{Correlation among the average variance from simulations (horizontal axis) 
and the variance from estimated (vertical axis) convergence maps for local and hemispherical analyses. 
The red open circles in the local analysis plot correspond to the regions which the estimated Planck convergence map deviate from the mean of the simulations more than $2\Sigma$, as identified by the $\chi^2$ analysis (see text for details)}.

\label{fig:correlation}
\end{figure}

We analyzed the statistical isotropy of the simulated and estimated Planck convergence maps using 
the variance as statistical estimator calculated in regions defined according to two approaches, 
small and large regions, as discussed in section \ref{sec:methodology}. 
As a result we observe that the variance from the data map is not isotropically distributed in the sky in none of the cases.
Moreover,
the comparison between the angular distribution of the variance on sky for the estimated 
convergence map and the average from simulated maps showed up a strong correlation in both, 
the local (figures \ref{fig:variance_plan} and \ref{fig:variance_sim}) and
hemispherical (figures \ref{fig:Hemispherical_planck} and \ref{fig:Hemispherical_simul}) 
analyses, as shown in the correlation plots in figure \ref{fig:correlation}. 

Such correlation would be an indication that the contributions coming from diverse 
non-lensing processes may have introduced the anisotropies imprinted in the variance 
distribution. An example is the Planck scanning strategy, which makes the instrumental noise to be minimised close to the ecliptic poles. Accordingly, the minimum variance values obtained from the local analysis of the convergence data, as well as those from the simulations (figures \ref{fig:variance_plan} and \ref{fig:variance_sim}), are displayed around the ecliptic poles. 
This corroborates the efficiency of the variance estimator, upon small and large regions, not only to probe the lensing signal, but also the residual contamination in the data and accounted in the simulations.

Even visually correlated, observing the color scales of the Mollweide projections in figures \ref{fig:variance_plan} and \ref{fig:variance_sim}, the variance in the estimated Planck map still appears to have some regions with higher/lower amplitudes compared to the average from 
the simulated maps. 
In order to evaluate the possibility of having regions of the $\hat{\kappa}^{WF}$ map with anomalously high (or low) variance values, we establish a comparison between its corresponding variance map and the one obtained from the simulations
through a $\chi^2$ analysis. 
Then, for each of the $i = h$th hemisphere, with $h = 1,\ldots, 192$, and $i = p$th patch, $p= 1\ldots,116$, we perform the calculation of \begin{equation}
\centering
\chi^2_i =\bigg(\frac{\sigma^2_{i (Planck)}-\langle{\sigma^2_{i(Sims)}\rangle}}{ \Sigma_{i}}\bigg)^2 , 
\label{eq:chi} 	
\end{equation}
where $\langle {\sigma^2_{i(Sims)}}\rangle$ and $\Sigma_{i}$ are the average and standard deviation over the variance sample of 100 simulated maps.

Associating to each region its corresponding $\chi^2$ value lead us to build a ${\chi}^2_p$-map for the local analysis and a ${\chi}^2_h$-map for the hemispherical one. 
These maps provide valuable informations about the contribution of the WL effect under different local 
regions and directions of the sky. 

The variance of the $\hat{\kappa}^{WF}$ and the mean variance of the simulated maps are in agreement on less than $1\sigma$ in the whole sky analysis- excluding only the portion of the WL mask as shown in the figure \ref{fig:histo_sky}.

\begin{figure}
	\hspace{-0cm}
	\vspace{-0cm}
	\includegraphics[width=0.45\textwidth]{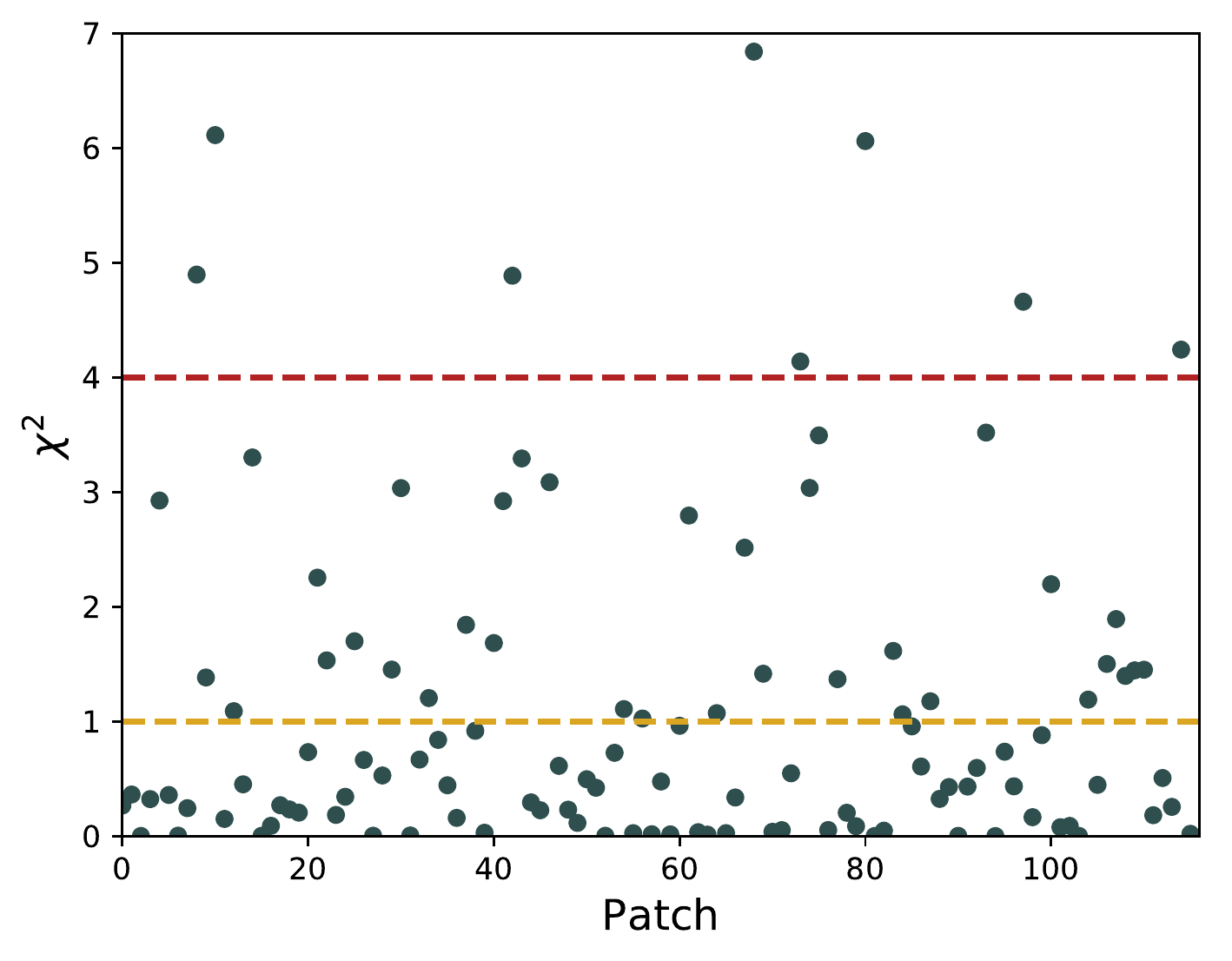} 
	\caption{$\chi^2_{p}$ values as a function of the patch number $p$, with $p = 1,\ldots, 116$. The dashed lines represent the corresponding 1$\Sigma$ (68\% CL; green) and 2$\Sigma$ (95\% CL; red) confidence regions, respectively.}
	\label{fig:qui_square_dis_patch}
\end{figure}

In fact, according to the $\chi^2_p$ distribution from the local analysis, showed in figure \ref{fig:qui_square_dis_patch}, most of the patches have 
a $\chi^2_p$ value very close to zero, indicating this agreement in those regions. For this reason, only the upper bounds of 68\% (yellow dashed line) and 95\% (red dashed line) Confidence Levels (CL) are shown in figure \ref{fig:qui_square_dis_patch}. However, we still find 8 patches whose ${\chi}^2_p$ values indicate that the variance of the data disagree in more than 2$\Sigma$ (i.e., 95\% CL) from the mean value of the simulations.

\begin{figure}
\hspace{-0cm}
\vspace{-0cm}
\includegraphics[width=0.45\textwidth]{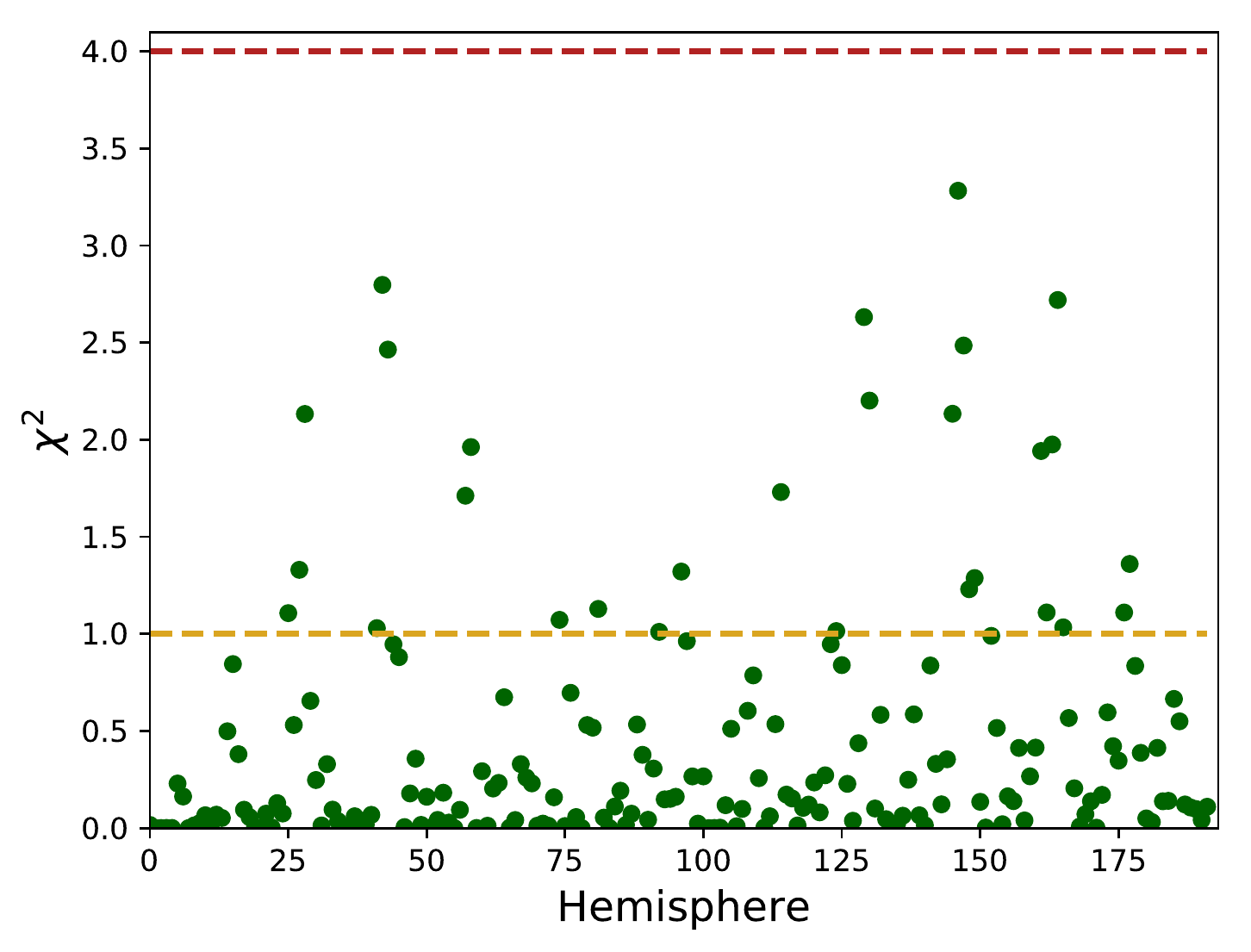} 
\caption{$\chi^2_{h}$ values as a function of the hemisphere number $h$, with $h = 1, ..., 192$. The dashed lines represent the $1\Sigma$ (68\% CL; yellow) and $2\Sigma$ (95\% CL; red) confidence regions, respectively.}
\label{fig:qui_square_dis_hemis}
\end{figure}

These 8 patches are represented as red open circles in the analysis shown in the upper panel of figure \ref{fig:correlation}. Note that the farther the symbols from the red diagonal line, the equality line ($\sigma^2_{p(Planck)} = \langle\sigma^2_{p(Sims)}\rangle$), higher the disagreement among the variance values from estimated and simulated convergence maps. The analysis presented in figure \ref{fig:qui_square_dis_patch} 
show how discrepant are the $\chi_p^2$ values of these 8 patches relatively to the others. In terms of the $\Sigma$ deviations of the variance from the mean of the simulations, 
the statistical significances of these patches, namely, 
$p = 68, 10, 80, 8, 42, 97, 114$\footnote{Notice that the numeration of the patches 
centres does not corresponds to the numeration of the hemispheres centres because 76 
patches were excluded by the WL mask.}, and $73$, are $2.61, 2.47, 2.46, 2.21, 2.21, 2.15, 2.05$, and $2.03~\Sigma$, respectively. Although do not seem to exist a statistically expressive discrepancy among data and simulations, these analyses identify outliers regions that may be associated to features of the lensing signal, or even some deficiency in the simulations.
The $\chi^2_{h}$ values for hemispherical analysis as a function of the hemisphere number are shown in figure \ref{fig:qui_square_dis_hemis}. The 1 and 2$\Sigma$ levels in this case are also represented in the figure as yellow and red dashed lines respectively. Similarly to the local analysis, most of the hemispheres of the estimated convergence map have their variance very close to those obtained from the simulations, so the corresponding $ \chi^2_{h}$ values are also quite small. In contrast to the local 
analysis results, the variance of the data upon all the hemispheres seems to be in 
agreement with the average from the simulations in less than $2\Sigma$. 
For this reason, there is no red open circles in the bottom panel of figure 
\ref{fig:correlation} as in the local results. 
However, it is worth mentioning that the hemispheres with the greatest deviations have their
variance values 1.84, 1.71, 1.68 and 1.66 $\Sigma$, whose coordinates centres are $h=146,42, 164$ and $129$, respectively.

\begin{figure}
\centering
\hspace{-0.3cm}
\includegraphics[scale=0.33]{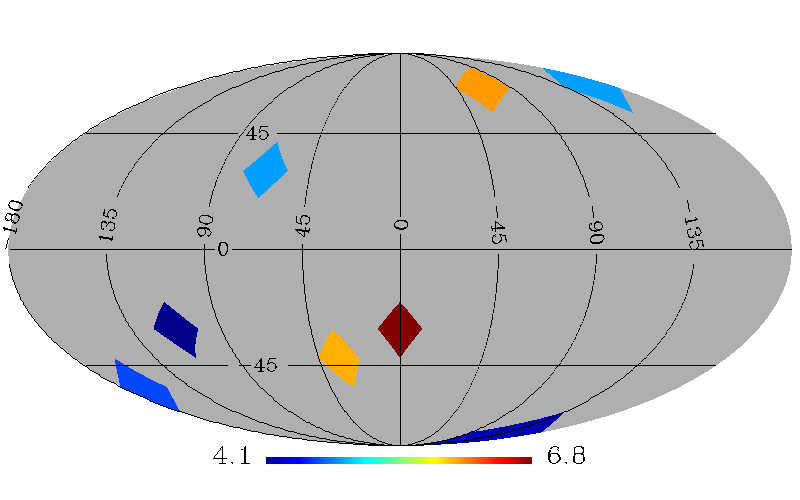} 
\vspace{-0.7cm}
\caption{Mollweide projection, in color scale, of the $\chi^2_p$-map, highlighting only the outliers patches, i.e., those whose $\chi^2_p$ values are out of the $2\Sigma$ confidence level (see 
figure \ref{fig:qui_square_dis_patch}).
}
\label{fig:highlighted_patches}
\end{figure}

The figure \ref{fig:highlighted_patches} shows, in color scale and Galactic coordinates, the 
sky projection of the outliers patches identified in the $\chi^2_{p}$-map. 
They do not appear to be concentrated in a particular region of the sky.  
However, one observes that some of these patches are located near the region defined by 
the WL mask, suggesting that residual foregrounds could still be affecting the analyses.

\begin{figure}
\hspace{-0.3cm}
\includegraphics[scale=0.33]{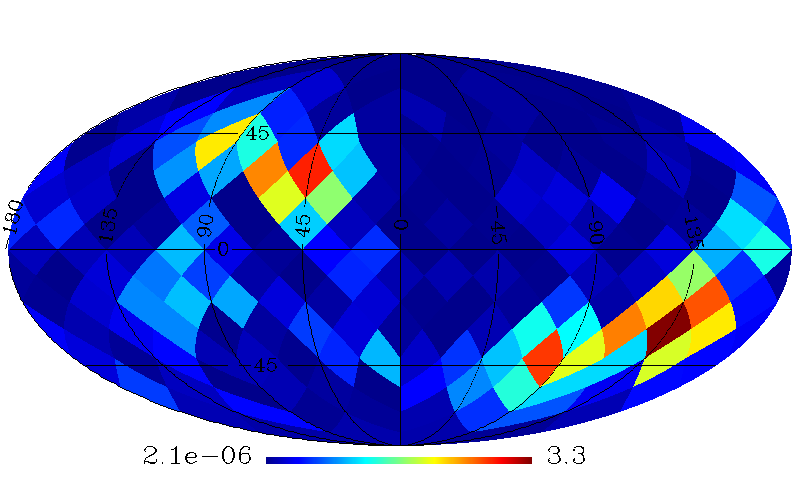} 
\vspace{-0.5cm}
	\caption{
	Mollweide projection, in color scale, of the $\chi^2_h$-map, for all of the 
	192 hemispheres centres (see figure \ref{fig:qui_square_dis_hemis}).
}
\label{fig:highlighted_hemis}
\end{figure}

Due to the different sizes encompassed in the local and hemispherical analyses, they probe different but complementary aspects. The hemispheres are composed by regions of $90\degree$ radius, probing effects dominant in large areas of the sky, while local analysis might reveal the effect of small structures or localized residual foregrounds. Unlike the local analysis, our results did not reveal outliers hemispheres, that is, where deviations are above 2$\Sigma$ level (95$\%$ CL). For this, we show in the figure \ref{fig:highlighted_hemis} the all sky projection of the $\chi^2_h$-map.

Although there are no significant deviations resulting from the hemispherical analysis, curiously, the regions with highest $\chi^2$, that is, the larger discrepancy among data and simulations, are concentrated around the ecliptic poles. Note that the hemispheres centred near the ecliptic poles are those where noise modelling could be the most problematic. This is expected because these hemispheres encompass regions where the noise levels are higher than average (i.e., near the galactic plane) and where deviations from statistical isotropy are expected due to the Planck scanning strategy.

The Planck convergence MV map used in our isotropy analyses is estimated 
from an optimized reconstruction using a quadratic estimator which measures the lensing 
induced in a CMB map, specifically, in the foreground-cleaned \texttt{SMICA} map 
\citep{adam2015planck}. 
However, as discussed by, e.g., \cite{ade2014planckisotro,novaes2016local,Schwarz2015cma}, 
the procedures used to obtain the four foreground-cleaned Planck CMB maps, among them the 
\texttt{SMICA}, even though highly efficient, could have leaved undesired residuals and 
some artifacts. 
Although the set of simulated maps take into account these effects, they might be under- or over-estimated. For example, the lack of accuracy of the statistically-isotropic and Gaussian field added to the simulations to take into account the noise in the data. In this sense, other sources of statistical anisotropy can also be the responsible for the small deviations of the variance observed in our analyses.

\begin{figure}
\hspace{-0cm}
\vspace{-0.cm}
\includegraphics[width=0.45\textwidth]{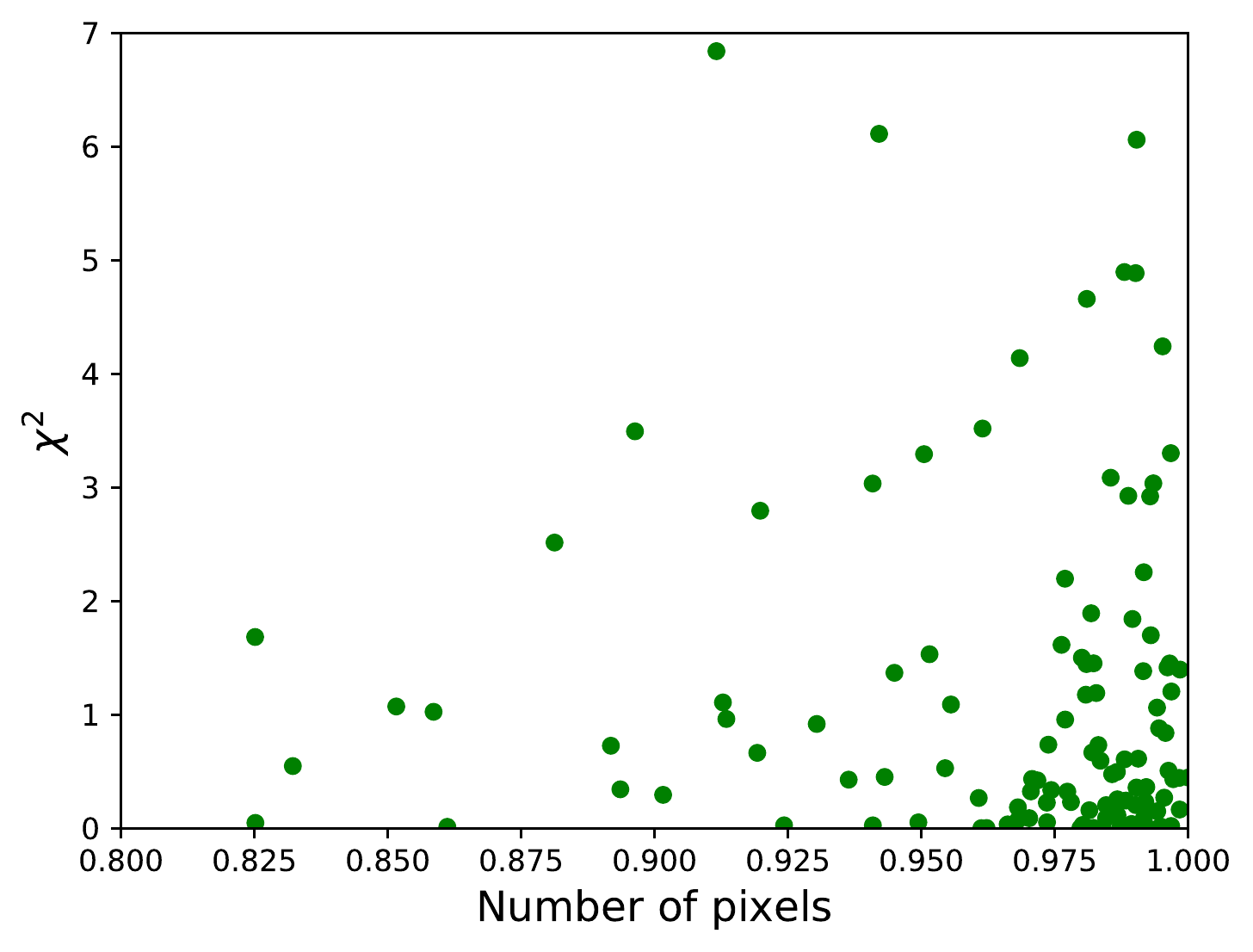} 
\caption{Dependence of the $\chi^2_p$ values with the percentage of valid pixels in the 116 patches. 
Notice that we are considering for analyses only the patches with a minimum of 80\% of valid pixels. 
}
\label{fig:pixels_patch}
\end{figure}

\begin{figure}
\hspace{-0cm}
\vspace{-0cm}
\includegraphics[width=0.45\textwidth]{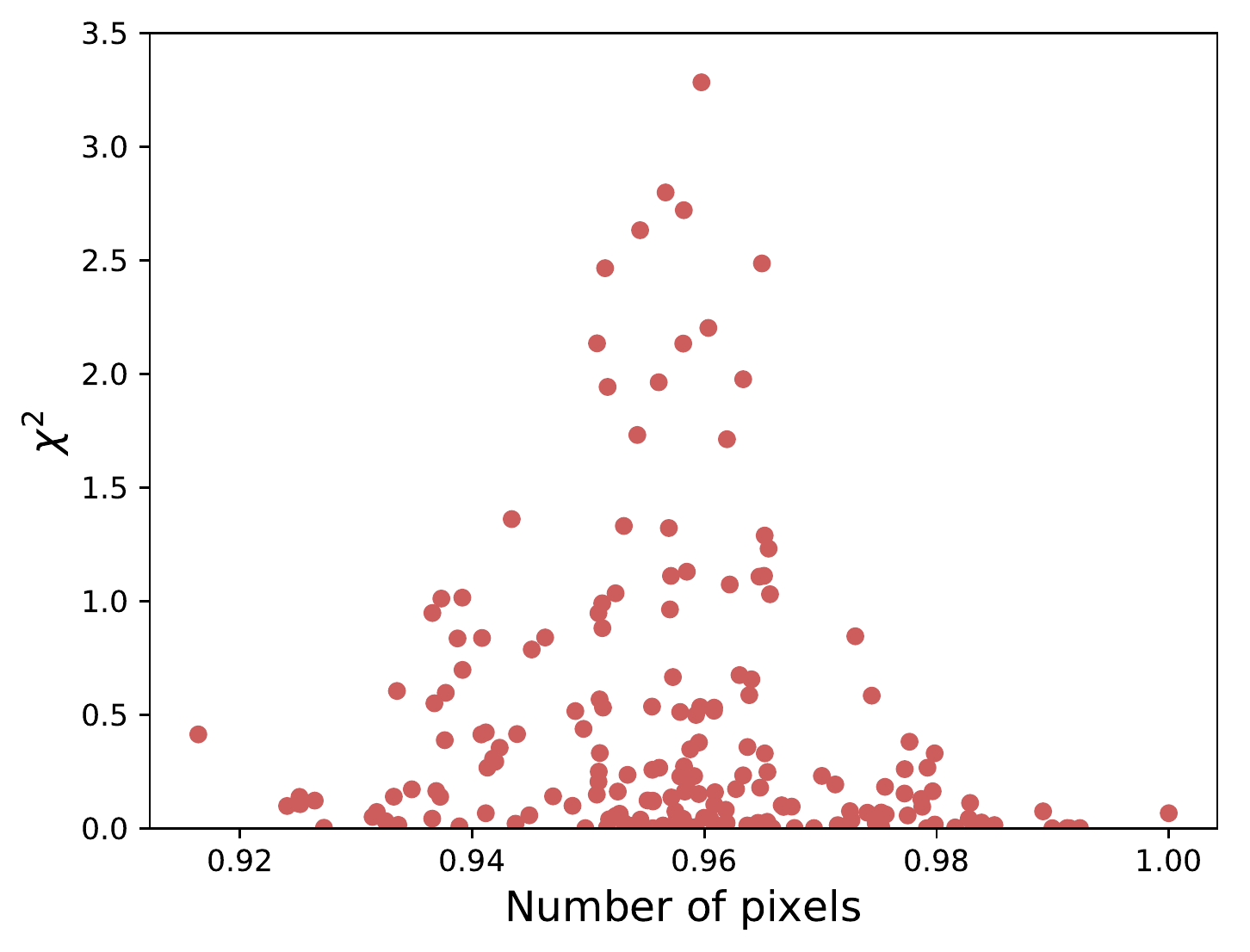} 
\caption{Dependence of the $\chi^2_{h}$ values with the percentage of valid pixels in the 192 hemispheres.}
\label{fig:pixels_hemis}
\end{figure}

Moreover, it is important to be sure that our results are not being influenced by the cut-sky 
mask applied to the convergence maps. 
As discussed before, all our analyses are performed considering a sky cut given by the WL mask, 
which means that each sky patch selected for the local analyses are composed by a 
different number of valid pixels, as well as in the case of the hemispherical analysis. 
In this case, to look for a possible influence of the mask in the local analysis, we analyse in 
figure \ref{fig:pixels_patch} the dependence between the $\chi^2_{p}$ values and the percentage 
of valid pixels for each of the 116 patches. In this figure one observes that do not seem to 
exist a correlation between these two quantities. 
In the case of the hemispherical analyses, the dependence of the $\chi^2_{h}$ with the number 
of valid pixels are shown in the figure \ref{fig:pixels_hemis}. 
Also, does not appear a clear dependency between these two quantities, confirming that all the 
information provided by the $\chi^2_{p}$-map and $\chi^2_{h}$-map would come essentially from 
the $\hat{\kappa}^{\rm WF}$.

\section{Conclusions and Final Remarks}
\label{sec:conclusion}

The WL is a secondary effect and despite its low amplitude, it modifies some features of the 
primary anisotropies of the CMB. 
The convergence map represents an integrated measurement of the total matter distribution in 
the observable universe, since it is directly related to the mass inhomogeneities along the 
line of sight \citep{Lewis,ade2015planck}. 
Accordingly, the study of the CMB lensing phenomenon can provide direct information about the 
large-scale distribution of matter and, by extension, about how isotropically it is distributed 
in the universe.

We have presented here statistical isotropy analyses of the CMB lensing phenomenon. 
In order to check if there are regions of the sky 
exceptionally influenced by this effect we use the amplitude of the variance of the convergence 
map estimated by the Planck team. 
To do this, we analysed the sky data comparing the variance value in the estimated Planck convergence 
map, $\hat{\kappa}^{MV}$, with the simulated set of convergence maps. We performed two types of complementary analyses, upon small and large regions, each one 
considering a different criterion to select the regions where to apply our estimator. 

In these analyses we calculated the variance in small patches and in hemispherical 
regions of the sky in two datasets, the estimated Wiener-filtered convergence map, 
$\hat{\kappa}^{WF}$, and in 100 simulated convergence Wiener-filtered maps, that mimic the 
same effects and properties of the $\hat{\kappa}^{WF}$ map ~\citep{ade2015planck,fullfocal}. 
Through a $\chi^2$ analyses our two approaches identify some regions of the estimated convergence map with variance larger than expected 
as compared with simulations. The obtained results lead us to our main conclusions: 

\begin{itemize}
\item The local analyses highlighted 8 of the 116 regions. They correspond to regions where the variance reveals to be higher, or smaller, than the expected from the same sky regions of the simulated maps, disagreeing in more than $2\Sigma$ (95\% CL). 
The distribution of these patches are spread in different parts of the sky, but two of them are near the Galactic WL mask, suggesting a possible contamination by foreground residuals. In other words, although do not exist a highly significant tension among the variance of the data and the mean from simulations, our analysis identified outliers regions that may be associated to some high/low lensing contribution.
\vspace{0.2cm}
\item The $\chi^2_h$-map obtained from the hemispherical analysis reveals that 
when evaluating large regions of the sky, the data seem to be in better agreement with 
the simulations, since no outlier region (i.e., discrepancy of the variance values in 
more than 2$\Sigma$) was identified in this case.
This indicates that the effects responsible for the outlier patches are in fact of local 
origin, having their signal diluted when analysing large regions. However, even though not statistically significant, it is worth to point out that the 
locations of the coordinate centres of the hemispheres with the highest $\chi_h^2$ values are coincidentally concentrated close to the north and south ecliptic poles.
  
\end{itemize}

The convergence map reveals a sum of the lens effect from near to distant structures. 
Generically, our results may be indicating a possible residual contamination in the CMB 
data, or even a true imprint left by anomalous density regions in the convergence map. 
In case it comes from the matter distribution of the large scale structure, further 
investigation of the signal, not only of its source but also the scale it appears, 
is crucial to evaluate the validity of the cosmological principle.
However, it is also possible that our findings are related to the presence of unaccounted 
contributions of Galactic and extragalactic signals in the Planck CMB data, or even coming 
from the reconstruction process of the convergence map, as previously discussed. 
Aware of this, the distribution observed in the $\chi^2_{p}$-map indicates either the 
presence of some local effect in the estimated Planck convergence map or some over- or 
under-estimated residual effect in the simulated dataset. Moreover, these effects are diluted when performing the hemispherical analysis, results that suggest the presence of a residual anisotropic contamination, which could be originated by a mis-modelling of the additional noise included in the simulations.

In this scenario, it is clear the importance of performing the types of complementary analyses shown here. 
First, as a test of the statistical isotropy of the universe using the CMB lensing data, a very 
powerful probe of the large-scale structure at high redshifts. 
In fact, according to our results this observable seems to be in good agreement with what 
is expected in the concordance $\Lambda {\rm CDM}$ model. Second, to explore how the noise is distributed in the lensing data to be taken into account in future simulations. Our estimator complements other analyses helping to identify possible additional signals, that can be corroborated in future analyses with deep large-scale structure tracers. This comparison between data and simulations at small and large regions are a 
complementary and independent tests of the procedures used by the Planck collaboration in their attempt to simulated the data, accounting for the noise, foregrounds, and systematics of the measurements.

In fact, several works use the estimated Planck convergence map in some small 
specific region of the sky in order to obtain cosmological information along with 
large-scale structure data \citep{kirk2016cross,singh2017cross,liu2015cross,hand2015first}. 
This turns clear the importance of confirming whether the regions identified in the present 
work can offer some extra cosmological information. 

Finally, as a matter of fact, our analyses have shown that even though the universe seems to 
be in agreement with the principle of statistical isotropy, with no apparent preferred 
direction, we still observe localized regions that deserve future detailed analyses.
The scrutiny of these regions can lead us not only to find a possible correlation with under- 
or over-densities in the universe, but also to a better comprehension of how additional sources (besides those considered in the set of simulated Planck's convergence maps) could be affecting cosmological analyses.

\section*{Acknowledgements}

The authors thank CAPES, CNPq for the  financial support. 
We acknowledge a CAPES PVE project (88881.064966/2014-01) within the {\em Science without 
Borders Program}. CPN is also supported by the DTI-PCI Programme of the Brazilian Ministry 
of Science, Technology and Innovation (MCTI). ISF thanks CNPq’s grant PDE(234529/2014-08), 
and also FAPDF.
The authors also acknowledge the HEALpix package for the derivation of the analyses 
performed in this work.
We would like to thank the anonymous referee for very useful comments and feedback on 
this paper.


\vspace{1.cm}
\begin{thebibliography}{199}	
\bibitem[\protect\citeauthoryear{Aghanim, Majumdar  \& Silk}{Aghanim
	et~al.}{2008}]{aghanim2008secondary}
Aghanim N.,  Majumdar S.,   Silk J.,  2008, Reports on Progress in Physics, 71,
066902

\bibitem[\protect\citeauthoryear{Alonso, Salvador, S{\'a}nchez, Bilicki,
	Garc{\'\i}a-Bellido  \& S{\'a}nchez}{Alonso
	et~al.}{2015}]{alonso2015homogeneity}
Alonso D.,  Salvador A.~I.,  S{\'a}nchez F.~J.,  Bilicki M.,
Garc{\'\i}a-Bellido J.,   S{\'a}nchez E.,  2015, Monthly Notices of the Royal
Astronomical Society, 449, 670

\bibitem[\protect\citeauthoryear{Bengaly, Bernui, Ferreira  \& Alcaniz}{Bengaly
	et~al.}{2016a}]{bengaly2015probing}
Bengaly C. A.~P.,  Bernui A.,  Ferreira I.~S.,   Alcaniz J.~S.,  2016a, Monthly
Notices of the Royal Astronomical Society

\bibitem[\protect\citeauthoryear{Bengaly, Bernui, Alcaniz, Xavier  \&
	Novaes}{Bengaly et~al.}{2016b}]{bengaly2017there}
Bengaly C.,  Bernui A.,  Alcaniz J.,  Xavier H.,   Novaes C.,  2016b, Monthly
Notices of the Royal Astronomical Society, 464, 768

\bibitem[\protect\citeauthoryear{Bernui, Ferreira  \& Wuensche}{Bernui
	et~al.}{2008}]{bernui2008large}
Bernui A.,  Ferreira I.,   Wuensche C.,  2008, The Astrophysical Journal, 673,
968

\bibitem[\protect\citeauthoryear{Bernui, Oliveira  \& Pereira}{Bernui
	et~al.}{2014}]{bernui2014north}
Bernui A.,  Oliveira A.,   Pereira T.,  2014, Journal of Cosmology and
Astroparticle Physics, 2014, 041

\bibitem[\protect\citeauthoryear{Blanchard \& Schneider}{Blanchard \&
	Schneider}{1987}]{blanchard1987gravitational}
Blanchard A.,  Schneider J.,  1987, Astronomy and Astrophysics, 184, 1

\bibitem[\protect\citeauthoryear{Bobin, Starck, Sureau  \& Fadili}{Bobin
	et~al.}{2012}]{bobin2012cmb}
Bobin J.,  Starck J.-L.,  Sureau F.,   Fadili J.,  2012, Advances in Astronomy,
2012

\bibitem[\protect\citeauthoryear{Cole \& Efstathiou}{Cole \&
	Efstathiou}{1989}]{cole1989gravitational}
Cole S.,  Efstathiou G.,  1989, Monthly Notices of the Royal Astronomical
Society, 239, 195

\bibitem[\protect\citeauthoryear{Das et~al.,}{Das
	et~al.}{2011}]{das2011detection}
Das S.,  et~al., 2011, Physical Review Letters, 107, 021301

\bibitem[\protect\citeauthoryear{Das et~al.,}{Das
	et~al.}{2014}]{das2014atacama}
Das S.,  et~al., 2014, Journal of Cosmology and Astroparticle Physics, 2014,
014

\bibitem[\protect\citeauthoryear{Durrer}{Durrer}{2008}]{durrer2001theory}
Durrer R.,  2008, The cosmic microwave background.
Vol. 401, Cambridge University Press Cambridge

\bibitem[\protect\citeauthoryear{Feng, Keating, Paar  \& Zahn}{Feng
	et~al.}{2012a}]{feng2012recon}
Feng C.,  Keating B.,  Paar H.~P.,   Zahn O.,  2012a, Physical Review D, 85,
043513

\bibitem[\protect\citeauthoryear{Feng, Aslanyan, Manohar, Keating, Paar  \&
	Zahn}{Feng et~al.}{2012b}]{feng2012mea}
Feng C.,  Aslanyan G.,  Manohar A.~V.,  Keating B.,  Paar H.~P.,   Zahn O.,
2012b, Physical Review D, 86, 063519

\bibitem[\protect\citeauthoryear{Ghosh, Kothari, Jain  \& Rath}{Ghosh
	et~al.}{2016}]{ghosh2016dipole}
Ghosh S.,  Kothari R.,  Jain P.,   Rath P.~K.,  2016, Journal of Cosmology and
Astroparticle Physics, 2016, 046

\bibitem[\protect\citeauthoryear{G{\'o}rski, Banday, Hivon  \&
	Wandelt}{G{\'o}rski et~al.}{2002}]{gorski2002hea}
G{\'o}rski K.~M.,  Banday A.,  Hivon E.,   Wandelt B.,  2002, in Astronomical
Data Analysis Software and Systems XI. p.~107

\bibitem[\protect\citeauthoryear{Hand et~al.,}{Hand
	et~al.}{2015}]{hand2015first}
Hand N.,  et~al., 2015, Physical Review D, 91, 062001

\bibitem[\protect\citeauthoryear{Hanson, Smith, Challinor  \& Liguori}{Hanson
	et~al.}{2009}]{hanson2009cmb}
Hanson D.,  Smith K.~M.,  Challinor A.,   Liguori M.,  2009, Physical Review D,
80, 083004

\bibitem[\protect\citeauthoryear{Hanson, Challinor  \& Lewis}{Hanson
	et~al.}{2010}]{hanson2010weak}
Hanson D.,  Challinor A.,   Lewis A.,  2010, General Relativity and
Gravitation, 42, 2197

\bibitem[\protect\citeauthoryear{Hirata, Padmanabhan, Seljak, Schlegel  \&
	Brinkmann}{Hirata et~al.}{2004}]{hirata2004cross}
Hirata C.~M.,  Padmanabhan N.,  Seljak U.,  Schlegel D.,   Brinkmann J.,  2004,
Physical Review D, 70, 103501

\bibitem[\protect\citeauthoryear{Hirata, Ho, Padmanabhan, Seljak  \&
	Bahcall}{Hirata et~al.}{2008}]{hirata2008co}
Hirata C.~M.,  Ho S.,  Padmanabhan N.,  Seljak U.,   Bahcall N.~A.,  2008,
Physical Review D, 78, 043520

\bibitem[\protect\citeauthoryear{Hu}{Hu}{2000}]{hu2000weak}
Hu W.,  2000, Physical Review D, 62, 043007

\bibitem[\protect\citeauthoryear{Keisler et~al.,}{Keisler
	et~al.}{2011}]{keisler2011me}
Keisler R.,  et~al., 2011, The Astrophysical Journal, 743, 28

\bibitem[\protect\citeauthoryear{Kirk et~al.,}{Kirk
	et~al.}{2016}]{kirk2016cross}
Kirk D.,  et~al., 2016, Monthly Notices of the Royal Astronomical Society, 459,
21

\bibitem[\protect\citeauthoryear{Kogut, Banday, Bennett, G{\'o}rski, Hinshaw,
	Smoot  \& Wright}{Kogut et~al.}{1996}]{kogut1996microwave}
Kogut A.,  Banday A.,  Bennett C.,  G{\'o}rski K.,  Hinshaw G.,  Smoot G.,
Wright E.,  1996, The Astrophysical Journal Letters, 464, L5

\bibitem[\protect\citeauthoryear{Lewis}{Lewis}{2005}]{2010/lewis}
Lewis A.,  2005, Physical Review D, 71, 083008

\bibitem[\protect\citeauthoryear{Lewis \& Challinor}{Lewis \&
	Challinor}{2006}]{Lewis}
Lewis A.,  Challinor A.,  2006, Physics Reports, 429, 1

\bibitem[\protect\citeauthoryear{Linder}{Linder}{1990}]{linder1990analysis}
Linder E.~V.,  1990, Monthly Notices of the Royal Astronomical Society, 243,
353

\bibitem[\protect\citeauthoryear{Liu \& Haiman}{Liu \&
	Haiman}{2016}]{liu2016origin}
Liu J.,  Haiman Z.,  2016, Physical Review D, 94, 043533

\bibitem[\protect\citeauthoryear{Liu \& Hill}{Liu \& Hill}{2015}]{liu2015cross}
Liu J.,  Hill J.~C.,  2015, Physical Review D, 92, 063517

\bibitem[\protect\citeauthoryear{Liu, Hill, Sherwin, Petri, B{\"o}hm  \&
	Haiman}{Liu et~al.}{2016}]{liu2016cmb}
Liu J.,  Hill J.~C.,  Sherwin B.~D.,  Petri A.,  B{\"o}hm V.,   Haiman Z.,
2016, Physical Review D, 94, 103501

\bibitem[\protect\citeauthoryear{Novaes, Bernui, Marques  \& Ferreira}{Novaes
	et~al.}{2016}]{novaes2016local}
Novaes C.,  Bernui A.,  Marques G.,   Ferreira I.,  2016, Monthly Notices of
the Royal Astronomical Society, 461, 1363

\bibitem[\protect\citeauthoryear{Okamoto \& Hu}{Okamoto \&
	Hu}{2003}]{okamoto2003cosmic}
Okamoto T.,  Hu W.,  2003, Physical Review D, 67, 083002

\bibitem[\protect\citeauthoryear{{Planck Collaboration} et~al.,}{{Planck
		Collaboration} et~al.}{2014}]{ade2014planck}
{Planck Collaboration} et~al., 2014, A\&A, 571, A17

\bibitem[\protect\citeauthoryear{{Planck Collaboration} et~al.,}{{Planck
		Collaboration} et~al.}{2016a}]{adam2015planck}
{Planck Collaboration} et~al., 2016a, A\&A, 594, A9

\bibitem[\protect\citeauthoryear{{Planck Collaboration} et~al.,}{{Planck
		Collaboration} et~al.}{2016b}]{fullfocal}
{Planck Collaboration} et~al., 2016b, Astronomy \& Astrophysics, 594, A12

\bibitem[\protect\citeauthoryear{{Planck Collaboration} et~al.,}{{Planck
		Collaboration} et~al.}{2016c}]{ade2015planck}
{Planck Collaboration} et~al., 2016c, A\&A, 594, A15

\bibitem[\protect\citeauthoryear{{Planck Collaboration} et~al.,}{{Planck
		Collaboration} et~al.}{2016d}]{ade2014planckisotro}
{Planck Collaboration} et~al., 2016d, A\&A, 594, A16

\bibitem[\protect\citeauthoryear{{Planck Collaboration} et~al.,}{{Planck
		Collaboration} et~al.}{2016e}]{bmode}
{Planck Collaboration} et~al., 2016e, Astronomy \& Astrophysics, 596, A102

\bibitem[\protect\citeauthoryear{Schwarz, Copi, Huterer  \& Starkman}{Schwarz
	et~al.}{2016}]{Schwarz2015cma}
Schwarz D.~J.,  Copi C.~J.,  Huterer D.,   Starkman G.~D.,  2016,
Class.Quant.Grav. 33, no. 18, 184001

\bibitem[\protect\citeauthoryear{Serra \& Cooray}{Serra \&
	Cooray}{2008}]{serra2008impact}
Serra P.,  Cooray A.,  2008, Physical Review D, 77, 107305

\bibitem[\protect\citeauthoryear{Singh, Mandelbaum  \& Brownstein}{Singh
	et~al.}{2017}]{singh2017cross}
Singh S.,  Mandelbaum R.,   Brownstein J.~R.,  2017, Monthly Notices of the
Royal Astronomical Society, 464, 2120

\bibitem[\protect\citeauthoryear{Smidt, Cooray, Amblard, Joudaki, Munshi,
	Santos  \& Serra}{Smidt et~al.}{2011}]{smidt2011constraint}
Smidt J.,  Cooray A.,  Amblard A.,  Joudaki S.,  Munshi D.,  Santos M.~G.,
Serra P.,  2011, The Astrophysical Journal Letters, 728, L1

\bibitem[\protect\citeauthoryear{Smith, Zahn  \& Dore}{Smith
	et~al.}{2007}]{smith2007d}
Smith K.~M.,  Zahn O.,   Dore O.,  2007, Physical Review D, 76, 043510

\bibitem[\protect\citeauthoryear{Story et~al.,}{Story
	et~al.}{2013}]{story2013measurement}
Story K.,  et~al., 2013, The Astrophysical Journal, 779, 86

\bibitem[\protect\citeauthoryear{Tarnopolski}{Tarnopolski}{2015}]{tarnopolski2015testing}
Tarnopolski M.,  2015, arXiv preprint arXiv:1512.02865

\bibitem[\protect\citeauthoryear{Tiwari \& Nusser}{Tiwari \&
	Nusser}{2016}]{tiwari2016revisiting}
Tiwari P.,  Nusser A.,  2016, Journal of Cosmology and Astroparticle Physics,
2016, 062

\bibitem[\protect\citeauthoryear{Ukwatta \& Wo{\'z}niak}{Ukwatta \&
	Wo{\'z}niak}{2016}]{ukwatta2016inve}
Ukwatta T.,  Wo{\'z}niak P.,  2016, Monthly Notices of the Royal Astronomical
Society, 455, 703

\bibitem[\protect\citeauthoryear{Van~Engelen et~al.,}{Van~Engelen
	et~al.}{2012}]{van2012measurement}
Van~Engelen A.,  et~al., 2012, The Astrophysical Journal, 756, 142

\bibitem[\protect\citeauthoryear{Weiland et~al.,}{Weiland
	et~al.}{2010}]{weiland2010seven}
Weiland J.,  et~al., 2010, arXiv preprint arXiv:1001.4731
\end{thebibliography}

\bsp	
\label{lastpage}
\end{document}